\documentclass[aps,showpacs,pra]{revtex4}
\usepackage{graphicx}
\usepackage{dcolumn}
\usepackage{bm}
\usepackage{epsfig,psfig,pstricks,pst-grad,fancybox,graphics}
\usepackage{latexsym}
\usepackage{amsmath}
\usepackage{amssymb}
\usepackage{enumerate}
\usepackage{bbm}
\usepackage[latin1]{inputenc}

\begin{document}

\title{BCS pairing in a trapped dipolar Fermi gase.}
\author{M.A. Baranov, \mbox{\L}. Dobrek, and M. Lewenstein}
\affiliation{Institut f\"ur Theoretische Physik, Universit\"at Hannover, D-30167
Hannover,Germany}
\date{\today }

\begin{abstract}
We present a detailed study of the BCS pairing transition in a trapped
polarized dipolar Fermi gas. In the case of a shallow nearly spherical trap,
we find the decrease of the transition temperature as a function of the trap
aspect ratio and predict the existence of the optimal trap geometry. The
latter corresponds to the highest critical temperature of the BCS transition
for a given number of particles. We also derive the phase diagram for an
ultracold trapped dipolar Fermi gases in the situation, where the trap
frequencies can be of the order of the critical temperature of the BCS
transition in the homogeneous case, and find the critical value of the
dipole-dipole interaction energy, below which the BCS transition ceases to
exist. The critical dipole strength is obtained as a function of the trap
aspect ratio. Alternatively, for a given dipole strength there is a critical
value of the trap anisotropy for the BCS state to appear. The order
parameter calculated at criticality, exhibits nover non-monotonic behavior
resulted from the combined effect of the confining potential and anisotropic
character of the interparticle dipole-dipole interation.
\end{abstract}

\maketitle

\section{\protect\bigskip Introduction}

One of the most challenging goals of modern atomic, molecular and optical
physics is to observe the superfluid (BCS) transition \cite{deGennes} in
trapped Fermi gases. The possibility of such transition for gases with
attractive short range interactions has been predicted in Refs. \cite%
{shortrangeBCS}, and has been a subject of very intensive experimental
investigations since then (for the latest experimental results see \cite%
{BCSBEC}). In typical experiments evaporative cooling is used to cool
fermions. However, since the Pauli principle forbids the $s$-wave scattering
for fermions in the same internal state, Fermi-Fermi \cite{fermi-fermi} or
Fermi-Bose \cite{fermi-bose} mixtures have to be used to assure collisional
thermalization of the gas. Such combination of evaporation and sympathetic
cooling allows to reach temperatures $T\simeq 0.1T_{F}$, where $T_{F}$ is
the Fermi temperature at which the gas exhibits quantum degeneracy.
Unfortunately, critical temperatures for the BCS transition, $T_{c}$, are
predicted to be much smaller than $T_{F}$. Nowadays, the standard way
employs a Feshbach resonance in order to increase the atomic scattering
length $a_{s}$\ to larger negative values. Such \textquotedblright resonance
superfluidity\textquotedblright\ should lead to superfluid transition at $%
T_{c}\approx 0.1T_{F}$\ \cite{resonanceSF}. In the most promising scenario,
one starts with a molecular condensate formed for $a_{s}\geq 0$ and modifies 
$a_{s}$ towards the negative values \cite{BCSBEC,gora}. Another way to
achieve the BCS regime is to use the cooling scheme that can overcome the
Pauli blocking, such as appropriately designed laser cooling \cite{lasercool}%
. Yet another promising route is to place the Fermi gas in an optical
lattice and enter the "high $T_{c}$" regime \cite{corrlattice}.

The temperature of the BCS transition in a two-component Fermi gas depends
dramatically on the difference of concentrations of the two components,
which presents another experimental obstacle \cite{gora}. This problem,
however, is not relevant for a polarized Fermi gas with long-range
interactions, such as dipole-dipole ones. The possibility of the Cooper
pairing has been predicted in Refs. \cite{dippairing} and the critical
temperature (including many-body corrections), as well as the order
parameter, have been obtained for a homogeneous gas in Ref. \cite{dipoleTc}.
Possible realizations of dipolar gases include ultracold heteronuclear
molecular gases \cite{Meijer}, atomic gases in a strong DC electric field 
\cite{YouMarinesku}, atomic gases with laser-induced dipoles \cite%
{laserinddip}, or with magnetic dipoles \cite{PfauDoyle}. For dipolar
moments $d$ of the order of one Debye and densities $n$ of $10^{12}\mathrm{cm%
}^{-3}$, $T_{c}$ should be in the range of $100\mathrm{nK}$, i.e.
experimentally feasible.

Dipole-dipole interaction is not only of long-range, but also anisotropic,
i.e. partially attractive and partially repulsive. Thus, the nature of the
interaction for trapped gases may be controlled by the geometry of the trap.
For a dipolar Bose gas in a cylindrical trap with the axial (radial)
frequency $\omega _{z}(\omega _{\rho })$, there exist a critical aspect
ratio $\lambda =(\omega _{z}/\omega _{\rho })^{1/2}$, above which the
Bose-condensed gas collapses if the atom number is too large \cite{Santos},
and below which the condensate exhibits the roton-maxon instability \cite%
{Santos-roton}. The trap geometry is also expected to control the physics of
trapped dipolar Fermi gases. So far, however, only partial results were
known \cite{Nobel}: analytic corrections to $T_{c}$\ in \textquotedblright
loose\textquotedblright\ traps, and solution of the case of an infinite
\textquotedblright slab\textquotedblright\ with $\omega _{\rho }=0$, and $%
\omega _{z}$\ finite. In the latter case there exists a critical frequency
above which the superfluid phase does not exist. Very recently, we have
reported results for the case of a general trap \cite{PRL}, and predicted
the critical dipole strength, below which the BCS pairing transition ceases
to exist.

In this paper we present the detailed derivation of the results of Refs. 
\cite{Nobel} and \cite{PRL}, and study the effect of the trap geometry on
the BCS transition in trapped dipolar Fermi gases. We first consider the
case of a shallow nearly spherical trap with the trap frequencis $\omega
_{z} $, $\omega _{\rho }$ much smaller than the critical temperature $T_{c}$
of a spatially homogeneous gas with the density equal to the maximal density
of a trapped gas sample, $\omega _{z},\omega _{\rho }\ll T_{c}$. In this
case, the presence of a confining potential results in a decrease of the
critical temperature as compared to the spatially homogeneous gas. In the
case of a strong confining potential, where $\omega _{z},\omega _{\rho }\sim
T_{c}$, we calculate the phase diagram in the plane $\Gamma -\lambda ^{-1}$,
where $\Gamma =36nd^{2}/\pi \mu $ is the strength of the dipole-dipole
interaction relative to the chemical potential $\mu $. Below the critical
value of the interaction, $\Gamma <\Gamma _{c}$, the BCS transition does not
take place. Similarly, for a given dipole interaction strength there is a
critical value of $\lambda ^{-1}$, above which the BCS state appears. We
determine the dependence of $\Gamma _{c}$ on $\lambda ^{-1}$, and calculate
the order parameter at the criticality. The order parameter exhibits a novel
non-monotonic behavior in strongly elongated cylindrical traps.

\section{BCS pairing in a dipolar Fermi gas}

We consider a dipolar Fermi gas polarized along the $z$-direction in a
cylindrical trap. The corresponding Hamiltonian reads%
\begin{eqnarray}
\widehat{H} &=&\int_{\mathbf{r}}\widehat{\psi }^{\dagger }(\mathbf{r})\left[
-\frac{\hbar ^{2}\nabla ^{2}}{2m}+U_{\mathrm{trap}}(\mathbf{r})-\mu \right] 
\widehat{\psi }(\mathbf{r})  \label{Hamiltonian} \\
&&+\frac{1}{2}\int_{\mathbf{r},\mathbf{r}^{\prime }}\widehat{\psi }^{\dagger
}(\mathbf{r})\widehat{\psi }^{\dagger }(\mathbf{r}^{\prime })V_{\mathrm{dip}%
}(\mathbf{r}-\mathbf{r}^{\prime })\widehat{\psi }(\mathbf{r}^{\prime })%
\widehat{\psi }(\mathbf{r}),  \notag
\end{eqnarray}%
where $\widehat{\psi }^{\dagger }(\mathbf{r})$ and $\widehat{\psi }(\mathbf{r%
})$ are the canonical fermionic creation and annihilitation operators of
particles with the mass $m$ and the dipolar moment $d$, $U_{\mathrm{trap}}(%
\mathbf{r})=m[\omega _{\rho }^{2}(x^{2}+y^{2})+\omega _{z}^{2}z^{2}]$ the
trapping potential, $\mu $ the chemical potential, and $V_{\mathrm{dip}}(%
\mathbf{r})=(d^{2}/r^{3})(1-3z^{2}/r^{2})$.

The BCS pairing is characterized by the order parameter $\Delta (\mathbf{r}%
_{1},\mathbf{r}_{2})=V_{d}(\mathbf{r}_{1}-\mathbf{r}_{2})\left\langle \hat{%
\Psi}(\mathbf{r}_{1})\hat{\Psi}(\mathbf{r}_{2})\right\rangle $, which
attains nonzero values below the critical temperature $T_{c}$. Just below $%
T_{c}$, the order parameter is a nontrivial solution of the BCS gap equation
(see. e.g. \cite{deGennes}) 
\begin{equation}
\Delta (\mathbf{r}_{1},\mathbf{r}_{2})=-V_{\mathrm{dip}}(\mathbf{r}_{1}-%
\mathbf{r}_{2})\int_{\mathbf{r}_{3},\mathbf{r}_{4}}K(\mathbf{r}_{1},\mathbf{r%
}_{2};\mathbf{r}_{3},\mathbf{r}_{4})\Delta (\mathbf{r}_{3},\mathbf{r}_{4})
\label{1}
\end{equation}%
and is an extremum of the functional 
\begin{equation}
F[\Delta ]=\int_{\mathbf{r}_{1},\mathbf{r}_{2}}\frac{\left\vert \Delta (%
\mathbf{r}_{1},\mathbf{r}_{2})\right\vert ^{2}}{V_{\mathrm{dip}}(\mathbf{r}%
_{1}-\mathbf{r}_{2})}+\int_{\mathbf{r}_{1}\ldots \mathbf{r}_{4}}\Delta
^{\ast }(\mathbf{r}_{1},\mathbf{r}_{2})K(\mathbf{r}_{1},\mathbf{r}_{2};%
\mathbf{r}_{3},\mathbf{r}_{4})\Delta (\mathbf{r}_{3},\mathbf{r}_{4}).
\label{2}
\end{equation}%
The kernel $K$ in the above expression is 
\begin{equation}
K(\mathbf{r}_{1},\mathbf{r}_{2};\mathbf{r}_{3},\mathbf{r}_{4})=T\sum_{\omega
_{n}}G_{\omega _{n}}(\mathbf{r}_{1},\mathbf{r}_{3})G_{-\omega _{n}}(\mathbf{r%
}_{2},\mathbf{r}_{4}),  \label{3}
\end{equation}%
where 
\begin{equation*}
G_{\omega _{n}}(\mathbf{r},\mathbf{r}^{\prime })=\sum_{\nu }\frac{\psi _{\nu
}(\mathbf{r})\psi _{\nu }^{\ast }(\mathbf{r}^{\prime })}{i\omega
_{n}-(\varepsilon _{\nu }-\mu )}
\end{equation*}%
is the Matsubara Green function of the Fermi gas in the normal phase. In
this formula, $\omega _{n}=\pi T(2n+1)$, $n=0,\pm 1,\pm 2,\ldots $, $\psi
_{\nu }(\mathbf{r})$ and $\varepsilon _{\nu }$ are the eigenfunctions and
the corresponding eigenvalues of the free particle Hamiltonian in the trap, $%
[-\hbar ^{2}\nabla ^{2}/2m+V_{\mathrm{trap}}(\mathbf{r})]\psi _{\nu }(%
\mathbf{r})=\varepsilon _{\nu }\psi _{\nu }(\mathbf{r})$.

The solution of the general problem given by Eqs. (\ref{1})--(\ref{3}), is
very difficult even for numerical methods due to the large number of
variables, relatively low symmetry of the system, and a long-range character
of the interparticle interaction. Remarkably, in the case of a shallow
nearly spherical trap, the solution can be found analytically, whereas the
general case can be treated by using a variational approach.

\section{Shallow nearly spherical trap}

We begin with the case of a weakly deformed spherical trap with the
frequencies, which are much less than critical temperature, $\omega _{\rho
},\omega _{z}\ll T_{c}$. In the new variables 
\begin{equation*}
\mathbf{R}=(\mathbf{r}_{1}+\mathbf{r}_{3})/2,\quad \mathbf{r}=\mathbf{r}_{1}-%
\mathbf{r}_{3},
\end{equation*}%
\begin{equation*}
\mathbf{R}^{\prime }=(\mathbf{r}_{2}+\mathbf{r}_{4})/2,\quad \mathbf{r}%
^{\prime }=\mathbf{r}_{2}-\mathbf{r}_{4},
\end{equation*}%
the second term in Eq. (\ref{2}) reads 
\begin{equation}
\int_{\mathbf{R},\mathbf{R}^{\prime },\mathbf{r},\mathbf{r}^{\prime }}\Delta
^{\ast }(\mathbf{R}+\mathbf{r}/2,\mathbf{R}^{\prime }+\mathbf{r}^{\prime
}/2)K(\mathbf{R},\mathbf{R}^{\prime },\mathbf{r},\mathbf{r}^{\prime })\Delta
(\mathbf{R}-\mathbf{r}/2,\mathbf{R}^{\prime }-\mathbf{r}^{\prime }/2),
\label{4}
\end{equation}%
where 
\begin{equation}
K(\mathbf{R},\mathbf{R}^{\prime },\mathbf{r},\mathbf{r}^{\prime
})=T\sum_{\omega _{n}}G_{\omega _{n}}(\mathbf{R},\mathbf{r})G_{-\omega _{n}}(%
\mathbf{R}^{\prime },\mathbf{r}^{\prime }).  \label{5}
\end{equation}

The kernel $K$ depends on variables $\mathbf{R}$ and $\mathbf{R}^{\prime }$
only due to the presence of the trapping potential, but, as a function of $%
\mathbf{r}$ and $\mathbf{r}^{\prime }$, the kernel decays rapidly for $%
r,r^{\prime }>\xi _{0}$, where $\xi _{0}=p_{F}/mT_{c}=v_{F}/T_{c}$ with $%
p_{F}=mv_{F}=\sqrt{2m\mu }$ being the Fermi momentum, determines the
characteristic scale for pairing correlations. Under the condition $\omega
_{\rho },\omega _{z}\ll T_{c}$, the gap $\Delta (\mathbf{r}_{1},\mathbf{r}%
_{2})=\Delta (\mathbf{R}_{12},\mathbf{r}_{12})$ is a slowly varying function
(on the scale $\xi _{0}$) of $\mathbf{R}_{12}=(\mathbf{r}_{1}+\mathbf{r}%
_{2})/2$ (see the end of this Section). At the same time, the Fourier
transform of $\Delta $ with respect to $\mathbf{r}_{12}=\mathbf{r}_{1}-%
\mathbf{r}_{2},$ 
\begin{equation}
\widetilde{\Delta }(\mathbf{R}_{12},\mathbf{p})=\int_{\mathbf{r}_{12}}\Delta
(\mathbf{R}_{12},\mathbf{r}_{12})\exp (-i\mathbf{pr}_{12}),  \label{6}
\end{equation}%
varies on a scale of the order of the Fermi momentum, $p\sim p_{F}\sim \hbar
n^{1/3}$, see Ref. \cite{dipoleTc}. It is therefore convenient to write Eq. (%
\ref{4}) in the form 
\begin{equation}
\int_{\mathbf{R}_{c}}\int_{\mathbf{r}_{c},\mathbf{r},\mathbf{r}^{\prime
}}\int_{\mathbf{P},\mathbf{q}}\widetilde{\Delta }^{\ast }(\mathbf{R}_{c}+%
\frac{\mathbf{r}+\mathbf{r}^{\prime }}{4},\mathbf{P}+\frac{\mathbf{q}}{2}%
)\exp (i\mathbf{qr}_{c}\mathbf{-}i\mathbf{P}(\mathbf{r}-\mathbf{r}^{\prime
}))K(\mathbf{R}_{c}+\frac{\mathbf{r}_{c}}{2},\mathbf{R}_{c}-\frac{\mathbf{r}%
_{c}}{2},\mathbf{r},\mathbf{r}^{\prime })\widetilde{\Delta }(\mathbf{R}_{c}-%
\frac{\mathbf{r}+\mathbf{r}^{\prime }}{4},\mathbf{P}-\frac{\mathbf{q}}{2}),
\label{7}
\end{equation}%
where $\mathbf{R}_{c}=(\mathbf{R}+\mathbf{R}^{\prime })/2=(\mathbf{R}_{12}+%
\mathbf{R}_{34})/2$ and $\mathbf{r}_{c}=\mathbf{R}-\mathbf{R}^{\prime }$,
and, keeping in mind that the pairing takes place in the central region of
the gas sample, where $U_{\mathrm{trap}}(\mathbf{R})\ll \mu $, we can expand
the order parameter in powers of $(\mathbf{r}+\mathbf{r}^{\prime })/4$.

The leading term of this expansion is 
\begin{equation*}
\int_{\mathbf{R}_{c}}\int_{\mathbf{r}_{c},\mathbf{r},\mathbf{r}^{\prime
}}\int_{\mathbf{P},\mathbf{q}}\widetilde{\Delta }^{\ast }(\mathbf{R}_{c},%
\mathbf{P}+\frac{\mathbf{q}}{2})\exp (i\mathbf{q\mathbf{r}_{c}-}i\mathbf{P}(%
\mathbf{r}-\mathbf{r}^{\prime }))T\sum_{\omega _{n}}G_{\omega _{n}}(\mathbf{R%
}_{c}+\frac{\mathbf{r}_{c}}{2},\mathbf{r})G_{-\omega _{n}}(\mathbf{R}_{c}-%
\frac{\mathbf{r}_{c}}{2},\mathbf{r}^{\prime })\widetilde{\Delta }(\mathbf{R}%
_{c},\mathbf{P}-\frac{\mathbf{q}}{2})
\end{equation*}%
\begin{equation}
=\int_{\mathbf{R}_{c}}\int_{\mathbf{P},\mathbf{q}}\int_{\mathbf{r}_{c}}%
\widetilde{\Delta }^{\ast }(\mathbf{R}_{c},\mathbf{P}+\frac{\mathbf{q}}{2}%
)\exp (i\mathbf{qr}_{c})T\sum_{\omega _{n}}G_{\omega _{n}}(\mathbf{R}_{c}+%
\frac{\mathbf{r}_{c}}{2},\mathbf{P})G_{-\omega _{n}}(\mathbf{R}_{c}-\frac{%
\mathbf{r}_{c}}{2},-\mathbf{P})\widetilde{\Delta }(\mathbf{R}_{c},\mathbf{P}-%
\frac{\mathbf{q}}{2}),  \label{8}
\end{equation}%
with 
\begin{equation*}
G_{\omega _{n}}(\mathbf{R},\mathbf{P})=\int_{\mathbf{r}}G_{\omega _{n}}(%
\mathbf{R},\mathbf{r})\exp (-i\mathbf{Pr}).
\end{equation*}

In the case $\omega _{\rho },\omega _{z}\ll T_{c}$, the Green function $%
G_{\omega _{n}}(\mathbf{R},\mathbf{P})$ can be approximated as 
\begin{equation}
G_{\omega _{n}}(\mathbf{R},\mathbf{P})\approx \frac{1}{i\omega _{n}-\xi
_{P}+U_{\mathrm{trap}}(\mathbf{R})}=\frac{1}{i\omega _{n}-(P^{2}/2m-\mu (%
\mathbf{R}))}  \label{9}
\end{equation}%
with $\mu (\mathbf{R})=\mu -U_{\mathrm{trap}}(\mathbf{R})$. With the help of
the formula 
\begin{equation*}
T\sum_{\omega _{n}}\frac{1}{i\omega _{n}-\xi _{P}+U_{\mathrm{trap}}(\mathbf{R%
})}\frac{1}{-i\omega _{n}-\xi _{P}+U_{\mathrm{trap}}(\mathbf{R}^{\prime })}%
\approx \left[ 1-\frac{U_{\mathrm{trap}}(\mathbf{R})+U_{\mathrm{trap}}(%
\mathbf{R}^{\prime })}{2\mu }\frac{\partial }{\partial \xi _{P}}\right] 
\frac{\tanh (\xi _{P}/2T_{c})}{2\xi _{P}}
\end{equation*}%
\begin{equation}
=\frac{\tanh (\xi _{P}/2T_{c})}{2\xi _{P}}-\frac{1}{\mu }(U_{\mathrm{trap}}(%
\mathbf{R}_{c})+U_{\mathrm{trap}}(\mathbf{r}_{c}/2))\frac{\partial }{%
\partial \xi _{P}}\frac{\tanh (\xi _{P}/2T_{c})}{2\xi _{P}},  \label{10}
\end{equation}%
the integration over $\mathbf{r}_{c}$ in Eq. (\ref{8}) gives 
\begin{equation*}
\int_{\mathbf{r}_{c}}\exp (i\mathbf{q\mathbf{r}_{c}\rho })T\sum_{\omega
_{n}}G_{\omega _{n}}(\mathbf{R}_{c}+\mathbf{r}_{c}/2,\mathbf{P})G_{-\omega
_{n}}(\mathbf{R}_{c}-\mathbf{r}_{c}/2,-\mathbf{P})
\end{equation*}%
\begin{equation}
\approx \left( 1-\frac{1}{\mu }U_{\mathrm{trap}}(\mathbf{R}_{c})\frac{%
\partial }{\partial \xi _{P}}\right) \frac{\tanh (\xi _{P}/2T_{c})}{2\xi _{P}%
}\delta (\mathbf{q})+\frac{1}{\mu }U_{\mathrm{trap}}(\frac{1}{2}\mathbf{%
\nabla }_{\mathbf{q}})\delta (\mathbf{q})\frac{\partial }{\partial \xi _{P}}%
\frac{\tanh (\xi _{P}/2T_{c})}{2\xi _{P}}.  \label{11}
\end{equation}%
The term containing $U_{\mathrm{trap}}(\mathbf{\nabla }_{\mathbf{q}}/2)$ can
be neglected because $\nabla _{\mathbf{q}}\sim 1/p_{F}$, and, therefore, Eq.
(\ref{8}) can finally be written in the form 
\begin{equation}
\int_{\mathbf{R}_{c}}\int_{\mathbf{P}}\widetilde{\Delta }^{\ast }(\mathbf{R}%
_{c},\mathbf{P})\left( 1-\frac{1}{\mu }U_{\mathrm{trap}}(\mathbf{R}_{c})%
\frac{\partial }{\partial \xi _{P}}\right) \frac{\tanh (\xi _{P}/2T_{c})}{%
2\xi _{P}}\widetilde{\Delta }(\mathbf{R}_{c},\mathbf{P}).  \label{12}
\end{equation}%
This expression corresponds to the local density approximation with $\mu
\rightarrow \mu (\mathbf{R}_{c})=\mu -U_{\mathrm{trap}}(\mathbf{R}_{c})$
expanded in powers of $U_{\mathrm{trap}}(\mathbf{R}_{c})/\mu $ up to the
first order.

The next term of the expansion of Eq. (\ref{7}) in powers of $(\mathbf{r}+%
\mathbf{r}^{\prime })/4$ is quadratic (the linear in $(\mathbf{r}+\mathbf{r}%
^{\prime })/4$ contribution vanishes due to the symmetry of the problem) and
has the form 
\begin{equation*}
\int_{\mathbf{R}_{c}}\int_{\mathbf{r}_{c}}\int_{\mathbf{P},\mathbf{q}}\left( 
\widetilde{\Delta }^{\ast }(\mathbf{R}_{c},\mathbf{P}+\mathbf{q}/2)\left[ 
\overleftarrow{\nabla }_{i}\overleftarrow{\nabla }_{j}+\overrightarrow{%
\nabla }_{i}\overrightarrow{\nabla }_{j}-2\overleftarrow{\nabla }_{i}%
\overrightarrow{\nabla }_{j}\right] \widetilde{\Delta }(\mathbf{R}_{c},%
\mathbf{P}-\mathbf{q}/2)\right) \exp (i\mathbf{q\mathbf{r}_{c})}
\end{equation*}%
\begin{equation}
\times \int_{\mathbf{r},\mathbf{r}^{\prime }}\frac{1}{2}\left( \frac{\mathbf{%
r}+\mathbf{r}^{\prime }}{4}\right) _{i}\left( \frac{\mathbf{r}+\mathbf{r}%
^{\prime }}{4}\right) _{j}\exp \mathbf{(-}i\mathbf{P}(\mathbf{r}-\mathbf{r}%
^{\prime }))K(\mathbf{R}_{c}+\mathbf{r}_{c}/2,\mathbf{R}_{c}-\mathbf{r}%
_{c}/2,\mathbf{r},\mathbf{r}^{\prime })  \label{13}
\end{equation}%
where $\overleftarrow{\nabla }_{i}$ and $\overrightarrow{\nabla }_{i}$ are
the $i$-th component of the gradient $\nabla _{\mathbf{R}_{c}}$ acting on
the left (on $\widetilde{\Delta }^{\ast }$) and on the right (on $\widetilde{%
\Delta }$), respectively. After neglecting the $q$-dependence of $\widetilde{%
\Delta }$, the integrations over $\mathbf{r}_{c}$ and $\mathbf{q}$ are
straightforward and Eq. (\ref{13}) can written as 
\begin{equation*}
\int_{\mathbf{R}_{c}}\int_{\mathbf{P}}\left( \widetilde{\Delta }^{\ast }(%
\mathbf{R}_{c},\mathbf{P})\left[ \overleftarrow{\nabla }_{i}\overleftarrow{%
\nabla }_{j}+\overrightarrow{\nabla }_{i}\overrightarrow{\nabla }_{j}-2%
\overleftarrow{\nabla }_{i}\overrightarrow{\nabla }_{j}\right] \widetilde{%
\Delta }(\mathbf{R}_{c},\mathbf{P})\right)
\end{equation*}%
\begin{equation}
\times \int_{\mathbf{r},\mathbf{r}^{\prime }}\frac{1}{2}\left( \frac{\mathbf{%
r}+\mathbf{r}^{\prime }}{4}\right) _{i}\left( \frac{\mathbf{r}+\mathbf{r}%
^{\prime }}{4}\right) _{j}\exp \mathbf{(-}i\mathbf{P}(\mathbf{r}-\mathbf{r}%
^{\prime }))K(\mathbf{R}_{c},\mathbf{R}_{c},\mathbf{r},\mathbf{r}^{\prime }).
\label{14}
\end{equation}%
One can show with the help of the explicit form of the Green functions, Eq. (%
\ref{9}), that the main contribution from the integrals over $\mathbf{r}$
and $\mathbf{r}^{\prime }$ is 
\begin{equation*}
\frac{1}{32}\mathbf{n}_{i}\mathbf{n}_{j}\frac{\mu (\mathbf{R}_{c})}{%
mT_{c}^{2}}\frac{1}{\cosh ^{2}(\xi _{P}(\mathbf{R}_{c})/2T_{c})}\frac{\tanh
(\xi _{P}(\mathbf{R}_{c})/2T_{c})}{2\xi _{P}(\mathbf{R}_{c})}
\end{equation*}%
with $\mathbf{n}$ being the unit vector in the direction of $\mathbf{P}$, \ $%
\mathbf{n}=\mathbf{P}/P$, and $\xi _{P}(\mathbf{R}_{c})=P^{2}/2m-\mu (%
\mathbf{R}_{c})$. This expression decays exponentially for $\left\vert \xi
_{P}(\mathbf{R}_{c})\right\vert >T_{c}$, and, therefore, can be approximated
by the simpler one in integrals over $\mathbf{P}$ with a slow varying
functions of $\mathbf{P}$ 
\begin{equation*}
\frac{1}{32}\mathbf{n}_{i}\mathbf{n}_{j}\frac{\mu (\mathbf{R}_{c})}{%
mT_{c}^{2}}\frac{7\zeta (3)}{2\pi ^{2}}\delta (P-p_{F}(\mathbf{R}_{c})),
\end{equation*}%
where $\zeta (z)$ is the Riemann zeta-function. With this simplification,
Eq. (\ref{14}) becomes 
\begin{equation}
\frac{7\zeta (3)}{64\pi ^{2}}\int_{\mathbf{R}_{c}}\nu _{F}(\mu (\mathbf{R}%
_{c}))\frac{\mu (\mathbf{R}_{c})}{mT_{c}^{2}}\int \frac{d\mathbf{n}}{4\pi }%
\left. \left( \widetilde{\Delta }^{\ast }(\mathbf{R}_{c},\mathbf{P})\left[ 
\overleftarrow{\nabla }_{i}\overleftarrow{\nabla }_{j}+\overrightarrow{%
\nabla }_{i}\overrightarrow{\nabla }_{j}-2\overleftarrow{\nabla }_{i}%
\overrightarrow{\nabla }_{j}\right] \widetilde{\Delta }(\mathbf{R}_{c},%
\mathbf{P})\right) \right\vert _{\mathbf{P}=\mathbf{n}p_{F}(\mathbf{R}_{c})},
\label{15}
\end{equation}%
where $\nu _{F}(\mu (\mathbf{R}_{c}))=mp_{F}(\mathbf{R}_{c})/2\pi ^{2}$ is
the density of states on the local Fermi surface.

After combining together Eqs. (\ref{12}) and (\ref{15}) and performing the
variation with respect to $\widetilde{\Delta }^{\ast }(\mathbf{R}_{c},%
\mathbf{P})$, we obtain the gap equation in the form 
\begin{eqnarray}
\widetilde{\Delta }(\mathbf{R}_{c},\mathbf{P}) &=&-\int_{\mathbf{P}^{\prime
}}V_{\mathrm{dip}}(\mathbf{P}-\mathbf{P}^{\prime })\frac{\tanh (\xi
_{P^{\prime }}(\mathbf{R}_{c})/2T_{c})}{2\xi _{P^{\prime }}(\mathbf{R}_{c})}%
\widetilde{\Delta }(\mathbf{R}_{c},\mathbf{P}^{\prime })  \notag \\
&&-\frac{7\zeta (3)}{16\pi ^{2}}\nu _{F}(\mu (\mathbf{R}_{c}))\int \frac{d%
\mathbf{n}^{\prime }}{4\pi }(\mathbf{n}^{\prime }\nabla _{\mathbf{R}%
_{c}})^{2}\left. \widetilde{\Delta }(\mathbf{R}_{c},\mathbf{P}^{\prime
})\right\vert _{\mathbf{P}^{\prime }=\mathbf{n}^{\prime }p_{F}(\mathbf{R}%
_{c})}.  \label{16}
\end{eqnarray}

The first line of Eq. (\ref{16}) coincides with the gap equation in the
spatially homogeneous case, see Ref.\cite{dipoleTc}, with $\mathbf{R}_{c}$
being a parameter. Therefore, following the method developed in Ref.\cite%
{dipoleTc}, we find that the order parameter on the local Fermi surface has
the form $\widetilde{\Delta }(\mathbf{R}_{c},\mathbf{P}=\mathbf{n}p_{F}(%
\mathbf{R}_{c}))=\widetilde{\Delta }(\mathbf{R}_{c})\varphi _{0}(\mathbf{n})$%
, where $\varphi _{0}(\mathbf{n})=\sqrt{2}\sin ((\pi /2)\cos (\vartheta _{%
\mathbf{n}}))$ (see Ref. \cite{dipoleTc} for more details) and the function $%
\widetilde{\Delta }(\mathbf{R}_{c})$ obeys the equation 
\begin{equation}
\widetilde{\Delta }(\mathbf{R}_{c})=-\lambda _{0}(\mathbf{R}_{c})\ln \frac{%
e^{C}\overline{\omega }(\mathbf{R}_{c})}{\pi T_{c}}\widetilde{\Delta }(%
\mathbf{R}_{c})-\frac{7\zeta (3)}{16\pi ^{2}}\lambda _{0}(0)\int \frac{d%
\mathbf{n}^{\prime }}{4\pi }\varphi _{0}^{2}(\mathbf{n}^{\prime })(\mathbf{n}%
^{\prime }\nabla _{\mathbf{R}_{c}})^{2}\widetilde{\Delta }(\mathbf{R}_{c}),
\label{17}
\end{equation}%
where $C=0.5772$ is the Euler constant, $\lambda _{0}(\mathbf{R})=(mp_{F}(%
\mathbf{R})/2\pi ^{2})\Gamma _{d}$, with $\Gamma _{d}$ being the
dipole-dipole scattering amplitude ($\Gamma _{d}=-8d^{2}/\pi $ in the Born
approximation). With the account of the expressions for the critical
temperature $T_{c}^{(0)}$ and explicite form of $\varphi _{0}(\mathbf{n})$
in the spatially homogeneous case, see Ref. \cite{dipoleTc}, the equation (%
\ref{17}) for $\widetilde{\Delta }(\mathbf{R}_{c})$ takes the final form 
\begin{equation}
\left\{ -\frac{7\zeta (3)}{48\pi ^{2}}\left( \frac{v_{F}}{T_{c}^{(0)}}%
\right) ^{2}\sum_{i=1,2,3}f_{i}\,\nabla _{\mathbf{R}_{c}i}^{2}+\frac{U_{%
\mathrm{trap}}(\mathbf{R}_{c})}{\mu }\left( 1-\frac{1}{2\lambda _{0}(0)}%
\right) \right\} \widetilde{\Delta }(\mathbf{R}_{c})=\ln \frac{T_{c}^{(0)}}{%
T_{c}}\widetilde{\Delta }(\mathbf{R}_{c}),  \label{18}
\end{equation}%
where $f_{1}=f_{2}=1-3/\pi ^{2}$, $f_{3}=1+6/\pi ^{2}$. Note, that in
obtaining this equation from Eq. (\ref{18}), we also expand $\lambda _{0}(%
\mathbf{R}_{c})$ up to the first order in $U_{\mathrm{trap}}(\mathbf{R}%
_{c})/\mu $, similarly to Eq. (\ref{12}).

Eq. (\ref{18}) is equivalent to the Schrödinger equation for a 3D
anisotropic harmonic oscillator. As a result, the shift in the critical
temperature due to the presence of the trapping potential is given by the
lowest eigenvalue and equals 
\begin{equation}
\frac{T_{c}^{(0)}-T_{c}}{T_{c}^{(0)}}\approx \ln \frac{T_{c}^{(0)}}{T_{c}}=%
\frac{1}{T_{c}^{(0)}}\sqrt{\frac{7\zeta (3)}{48\pi ^{2}}\left( 1+\frac{1}{%
2\left\vert \lambda _{0}\right\vert }\right) }\left[ 2\omega _{\rho }\sqrt{1-%
\frac{3}{\pi ^{2}}}+\omega _{z}\sqrt{1+\frac{6}{\pi ^{2}}}\right] .
\label{trapTc}
\end{equation}%
In the considered case $\omega _{i}/T_{c}\ll 1$, the critical temperature in
the trap is only slightly lower than that in the spatially homogeneous case.

Just below $T_{c}$, the order parameter has the gaussian form 
\begin{equation*}
\widetilde{\Delta }(\mathbf{R}_{c})\propto \exp
(-\sum_{i=1,2,3}R_{ci}^{2}/2l_{\Delta i}^{2}),
\end{equation*}%
where $l_{\Delta i}=(v_{F}/\omega _{i})\sqrt{\omega _{i}/T_{c}^{(0)}}\left[
7\zeta (3)f_{i}/48\pi ^{2}(1+1/2\left\vert \lambda _{0}\right\vert )\right]
^{1/4}$ is the characteristic size in the $i$-th direction. One can see,
that $l_{\Delta i}\ll R_{TF}^{(i)}$, where $R_{TF}^{(i)}=v_{F}/\omega _{i}$
is the Thomas-Fermi radius of the gas sample in the $i$-th direction. This
justifies our assumption that pairing takes place only in the central part
of the sample. On the other hand, we have $l_{\Delta i}\gg \xi _{0}$ and,
therefore, the gradient expansion of the order parameter in powers of $%
\mathbf{r}+\mathbf{r}^{\prime }$ is legitimate.

For a given number of particle $N=\mu ^{3}/3\omega _{z}\omega _{\rho }^{2}$
in the trap with aspect ratio $\lambda =\sqrt{\omega _{z}/\omega _{\rho }}$,
we have $\omega _{z}=\omega \lambda ^{4/3}$ and $\omega _{\rho }=\omega
\lambda ^{-2/3}$, where $\omega =(\omega _{z}\omega _{\rho }^{2})^{1/3}$,
and Eq. (\ref{trapTc}) becomes

\begin{equation}
\frac{T-T_{c}^{(0)}}{T_{c}^{(0)}}=-\frac{\omega }{T_{c}^{(0)}}\sqrt{\left( 1+%
\frac{1}{2\left\vert \lambda _{0}\right\vert }\right) }F(\lambda ),
\label{Tctrapaspr1}
\end{equation}%
where $F(\lambda )=\sqrt{7\zeta (3)/48\pi ^{2}}[2\sqrt{1-3/\pi ^{2}}\lambda
^{-2/3}+\sqrt{1+6/\pi ^{2}}\lambda ^{4/3}]$. The plot of the function $%
F(\lambda )$ is presented on Fig. 1.

\begin{figure}[tbp]
\epsfxsize 6.5cm \centerline{
\epsfbox{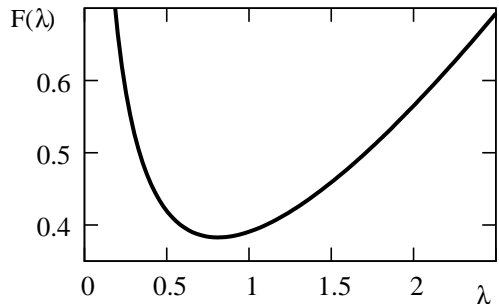}}
\caption{The function $F(\protect\lambda )$ versus the trap aspect ratio $%
\protect\lambda $.}
\end{figure}

We see that there exists the optimal value of the trap aspect ration $%
\lambda ^{\ast }=0.81$, which corresponds to the highest transition
temperature in the trap. The existence of the optimal value for the trap
aspect ratio is a result of the competition between the anisotropic
character of the dipole-dipole interparticles interaction and the
confinement of the gas sample in the radial direction. The former becomes
predominantly attractive for smaller values of $\lambda $ (cylindrical
traps) and, therefore, favours the BCS pairing. The latter, on the contrary,
acts on the pairing destructively due to the size effect and, therefore,
less pronounced for larger values of $\lambda $.

\section{Critical aspect ratio}

We have seen in the previous section, that the trap geometry has a more
pronounced influence on the BCS pairing in a dipolar Fermi gas, as compared
to a two-component Fermi gas with a short-range interaction. This is due to
the fact that the states, which form Cooper pairs in a trapped dipolar Fermi
gas, have different quantum numbers $n_{z}$. Therefore, their energies are
different, at least by the amount of $\omega _{z}$. When this difference
becomes of the order of $T_{c}$, the pairing is obviously impossible. As a
result, the superfluid transition in a trapped dipolar Fermi gas can take
place only in traps with $\omega _{z}<\omega _{zc}$, where the critical
frequency $\omega _{zc}$ is found to be $\omega _{zc}=1.8T_{c}$ \cite{Nobel}%
. For $\omega _{z}<\omega _{zc}$, as can be seen from Eq.(\ref{Tctrapaspr1}%
), the confinement in the radial direction decreases the critical
temperature as well. Therefore, one would expect the existence of the
critical aspect ratio $\lambda _{c}$ such that the pairing is possible only
in traps with $\lambda <\lambda _{c}$ \cite{PRL}.

In this Section we study the BCS pairing in the case of a cylindrical trap
where the trap frequencies can be of the order of the critical temperature
in the spatially homogeneous gas, $\omega _{\rho },\omega _{z}\sim
T_{c}^{(0)}$, but still much less than the chemical potential, $\omega
_{\rho },\omega _{z}\ll \mu $. In this case, the BCS gap equation (\ref{1})
can hardly be tractable even numerical without additional simplifications.

As it was shown in Ref. \cite{dipoleTc}, the BCS pairing is dominated by the 
$p$-wave scattering with zero projection of the angular momentum on the $z$%
-axis, $l_{z}=0$, in which the interaction is attractive. Contributions of
higher angular momenta, although present due to the long-range character of
the dipole-dipole interaction, are numerically small (see also Ref. \cite%
{dippairing}). In the $p$-wave channels with $l_{z}=\pm 1$ the interaction
is repulsive and, therefore, leads only to renormalizations of a
Fermi-liquid type, and will be neglected here. Therefore, for the considered
pairing problem we can model the dipole-dipole interaction by the following
(off-shell) scattering amplitude 
\begin{equation}
\Gamma _{1}(\mathbf{p},\mathbf{p}^{\prime },E)=p_{z}p_{z}^{\prime }%
\widetilde{\gamma }_{1}(E),  \label{gammadef}
\end{equation}%
where $\mathbf{p}$ is the incoming momentum, $\mathbf{p}^{\prime }$ the
outgoing one, and $\widetilde{\gamma }_{1}(E)$ some function of the energy $%
E $. The amplitude $\Gamma _{1}$ describes anisotropic scattering only\ in
the $p$-wave channel with the projection of the angular momentum $l_{z}=0$.
The function $\widetilde{\gamma }_{1}(E)$ obeys the equation 
\begin{equation}
\widetilde{\gamma }_{1}(E)-\widetilde{\gamma }_{1}(E^{\prime
})=\int^{\Lambda }\frac{d\mathbf{p}}{(2\pi )^{3}}\widetilde{\gamma }%
_{1}(E)\left( \frac{p_{z}^{2}}{p^{2}-E+i0}-\frac{p_{z}^{2}}{p^{2}-E^{\prime
}+i0}\right) \widetilde{\gamma }_{1}(E^{\prime }),  \label{gammaeq}
\end{equation}%
that follows from the Lipmann-Schwinger equation for the off-shell
scattering amplitude \cite{taylor}; $\widetilde{\gamma }_{1}(E)$ is assumed
to be negative in order to guarantee the BCS pairing. The cut-off parameter $%
\Lambda $ ensures the convergence of the integral and, in fact, can be
expressed in terms of the observable scattering data corresponding to
on-shell scattering amplitude with $p=p^{\prime }$ and $E=p^{2}/m$. It
follows from Eq. (\ref{gammaeq}) that $\widetilde{\gamma }_{1}(E)\,$is
inversely proportional to $E$, $\widetilde{\gamma }_{1}(E)=\gamma _{1}\
(2mE)^{-1}$, with some coefficient $\gamma _{1}$. Therefore, the on-shell
amplitude is energy independent, as it should be for low-energy scattering
on the dipole-dipole potential (see Ref. \cite{lowEdipscattering}).

The coefficient $\gamma _{1}$ determines the value of the critical
temperature $T_{c}$ of the BCS transition in a spatially homogeneous gas and
can be expressed through the dipole moment $d$ using the results of Ref. 
\cite{dipoleTc}. In a homogeneous gas, the order parameter has the form $%
\Delta (\mathbf{p})=p_{z}\Delta _{0}$ with some constant $\Delta _{0}$, and
the linearized gap equation implies

\begin{equation}
\frac{1}{\widetilde{\gamma }_{1}(\mu )}=-\int \frac{d\mathbf{p}}{(2\pi )^{3}}%
\frac{p_{z}^{2}}{2\xi _{p}}\left[ \tanh \frac{\xi _{p}}{2T_{c}}-1\right] ,
\label{gaphom}
\end{equation}%
where $\xi _{p}=p^{2}/2m-\mu $, and the bare interaction is renormalized in
terms of the scattering amplitude with $\widetilde{\gamma }_{1}(\mu )=\gamma
_{1}/p_{F}^{2}$ at the Fermi energy $\varepsilon _{F}=\mu =p_{F}^{2}/2m$
along the lines of Ref. \cite{GorkovMelikB} ($p_{F}$ is the Fermi momentum).
After integrating over $p$, we obtain the equation on $T_{c}$:

\begin{equation}
1=\frac{1}{3}\left\vert \gamma _{1}\right\vert \nu _{F}\left[ \ln \frac{2\mu 
}{T_{c}}-\frac{8}{3}-\ln \frac{\pi }{4}+C\right] ,  \label{Tchom}
\end{equation}%
where $\nu _{F}=mp_{F}/2\pi ^{2}$ is the density of states at the Fermi
energy. After comparing the solution of Eq. (\ref{Tchom}) with the result of
Ref. \cite{dipoleTc} for $T_{c}$, we find $\gamma _{1}=-24d^{2}/\pi $.

In the ordinary space, the scattering amplitude $\Gamma _{1}$is

\begin{equation}
\Gamma _{1}(\mathbf{r},\mathbf{r}^{\prime },E)=\partial _{z}\delta (\mathbf{r%
})\partial _{z^{\prime }}\delta (\mathbf{r}^{\prime })\widetilde{\gamma }%
_{1}(E),  \label{gammaspace}
\end{equation}%
where $\mathbf{r}$ and $\mathbf{r}^{\prime }$ are the relative distances
between the two incoming and the two outgoing particles, respectively.
Therefore, the order parameter in the trapped gas, $\Delta (\mathbf{r}_{1},%
\mathbf{r}_{2})\sim \left\langle \psi (\mathbf{r}_{1})\psi (\mathbf{r}%
_{2})\right\rangle $, has the form

\begin{equation}
\Delta (\mathbf{r}_{1},\mathbf{r}_{2})=\partial _{z}\delta (\mathbf{r}%
)\Delta _{0}(\mathbf{R}),  \label{deltaspace}
\end{equation}%
where $\mathbf{r}=\mathbf{r}_{1}-\mathbf{r}_{2}$ and $\mathbf{R}=(\mathbf{r}%
_{1}+\mathbf{r}_{2})/2$, and the corresponding equation for the critical
temperature is

\begin{eqnarray}
&&\frac{\Delta _{0}(\mathbf{R})}{\widetilde{\gamma }_{1}(\mu )}=-\int_{%
\mathbf{R}^{\prime }}\left[ \sum_{\mathbf{n}_{1},\mathbf{n}_{2}}M_{\mathbf{n}%
_{1}\mathbf{n}_{2}}(\mathbf{R})M_{\mathbf{n}_{1}\mathbf{n}_{2}}(\mathbf{R}%
^{\prime })\frac{\tanh \left( \xi _{1}/2T\right) +\tanh \left( \xi
_{2}/2T\right) }{2(\xi _{1}+\xi _{2})}\right.  \notag \\
&&\left. -\int \frac{d\mathbf{p}}{(2\pi )^{3}}\int \frac{d\mathbf{q}}{(2\pi
)^{3}}\frac{p_{z}^{2}}{2\xi _{p}+q^{2}/4m}\exp (i\mathbf{q}(\mathbf{R}-%
\mathbf{R}^{\prime }))\right] \Delta _{0}(\mathbf{R}^{\prime }).
\label{gapeqren}
\end{eqnarray}%
Here $\xi _{i}=\xi (\mathbf{n}_{i})$, $\mathbf{n}=(n_{x},n_{z},n_{z})$ are
the harmonic oscillator quantum numbers, $\xi (\mathbf{n})=\hbar \left[
\omega _{z}(n_{z}+1/2)+\omega _{\rho }(n_{x}+n_{y}+1)\right] -\mu $, and the
function $M_{n_{1}n_{2}}(\mathbf{R})$ is defined as

\begin{equation}
M_{\mathbf{n}_{1}\mathbf{n}_{2}}(\mathbf{R}%
)=M_{n_{1z}n_{2z}}^{(z)}(z)M_{n_{1x}n_{2x}}^{(\rho
)}(x)M_{n_{1y}n_{2y}}^{(\rho )}(y)  \label{M}
\end{equation}
with

\begin{equation}
M_{n_{1}n_{2}}^{(z)}(z)=\frac{1}{2}\left[ \varphi _{n_{1}}(z)\partial
_{z}\varphi _{n_{2}}(z)-\varphi _{n_{2}}(z)\partial _{z}\varphi _{n_{1}}(z)%
\right] ,\quad M_{n_{1}n_{2}}^{(\rho )}(x)=\varphi _{n_{1}}(x)\varphi
_{n_{2}}(y),  \label{MzMro}
\end{equation}%
$\varphi _{n}(z)$ being the harmonic oscillator wave functions.

The gap equation (\ref{gapeqren}) is still hardly tractable numerically and,
hence, we employ additional approximations. We assume a large number of
particles such that the chemical potential $\mu $ is much larger than the
trap frequencies, $\mu \gg \omega _{z},\omega _{\rho }$. Therefore, while
calculating the functions $M_{\mathbf{n}_{1}\mathbf{n}_{2}}(\mathbf{R})$, we
can use the WKB approximation for the wave functions $\varphi _{n}$ of the
most important for the BCS pairing states with energies near the Fermi
energy $\varepsilon _{F}=\mu $. Another simplification is due to the fact
that the BCS order parameter $\Delta _{0}(\mathbf{R})$ varies slowly on an
interparticle distance scale $n^{-1/3}\sim \hbar /p_{F}$, where $p_{F}=\sqrt{%
2m\mu }$ is now the Fermi momentum in the center of the trap in the
Thomas-Fermi approximation. As a result, the pairing correlations are
pronounced only between states that are close in energy. This allows to
neglect $q^{2}/4m$ in the denominator of the second term in Eq.(\ref%
{gapeqren}) together with rapidly oscillating terms in the functions $M_{%
\mathbf{n}_{1}\mathbf{n}_{2}}(\mathbf{R})$. We then obtain (see Appendix)

\begin{equation}
M_{n_{1}n_{2}}^{(z)}(z)\equiv M_{Nn}^{(z)}(z)\approx (-1)^{n}\frac{m\omega
_{z}}{\pi }\sqrt{1-(z/l_{zN})^{2}}\mathrm{U}_{n-1}(z/l_{zN})  \label{WKBMz}
\end{equation}
and

\begin{equation}
M_{n_{1}n_{2}}^{(\rho )}(x)\equiv M_{Nn}^{(\rho )}(x)\approx (-1)^{n}\frac{1%
}{\pi l_{\rho N}}\frac{1}{\sqrt{1-(x/l_{\rho N})^{2}}}\mathrm{T}%
_{n}(x/l_{\rho N}),  \label{WKBMro}
\end{equation}%
where $n\equiv \left\vert n_{1}-n_{2}\right\vert \ll N\equiv (n_{1}+n_{2})/2$%
, $l_{iN}=\sqrt{2N\hbar /n\omega _{i}}=l_{0i}\sqrt{2N}$, whereas $\mathrm{U}%
_{n}(z)=\sin ((n+1)\arccos z)/\sin (\arccos z)$ and $\mathrm{T}_{n}(x)=\cos
(n\arccos x)$ are the Chebyshev polynomials. The functions $%
M_{n_{1}n_{2}}^{(z)}(z)$ and $M_{n_{1}n_{2}}^{(\rho )}(x)$ fullfil the
following completeness relations

\begin{eqnarray}
\sum\limits_{n\geq 1}M_{Nn}^{(z)}(z)M_{Nn}^{(z)}(z^{\prime }) &=&\frac{%
(n\omega _{z})^{2}}{2\pi }\sqrt{l_{zN}^{2}-z^{2}}\delta (z-z^{\prime })
\label{sumcompltz} \\
\sum\limits_{n\geq 0}\delta _{n}M_{Nn}^{(\rho )}(x)M_{Nn}^{(\rho
)}(x^{\prime }) &=&\frac{1}{\pi }\frac{1}{\sqrt{l_{\rho N}^{2}-x^{2}}}\delta
(x-x^{\prime }),  \label{sumcompltro}
\end{eqnarray}%
with $\delta _{0}=1$ and $\delta _{n>0}=2$, which follow from the
completeness of the Chebyshev polynomials.

It is convenient to rewrite Eq. (\ref{gapeqren}) in the following way

\begin{eqnarray}
-\frac{\Delta _{0}(\mathbf{R})}{\widetilde{\gamma }_{1}(\mu )} &\approx
&\int d\mathbf{R}^{\prime }\left[ \sum_{\mathbf{n}_{1},\mathbf{n}_{2}}M_{%
\mathbf{n}_{1}\mathbf{n}_{2}}(\mathbf{R})\left\{ \frac{\tanh \left( \xi
_{1}/2T\right) +\tanh \left( \xi _{2}/2T\right) }{2(\xi _{1}+\xi _{2})}-%
\frac{\tanh ((\xi _{1}+\xi _{2})4T)}{\xi _{1}+\xi _{2}}\right\} M_{\mathbf{n}%
_{1}\mathbf{n}_{2}}(\mathbf{R}^{\prime })\right.  \notag \\
&&+\sum_{\mathbf{n}_{1},\mathbf{n}_{2}}M_{\mathbf{n}_{1}\mathbf{n}_{2}}(%
\mathbf{R})\left\{ \frac{\tanh ((\xi _{1}+\xi _{2})4T)}{2(\xi _{1}+\xi _{2})}%
-\frac{1}{\xi _{1}+\xi _{2}}\right\} M_{\mathbf{n}_{1}\mathbf{n}_{2}}(%
\mathbf{R}^{\prime })  \notag \\
&&\left. \sum_{\mathbf{n}_{1},\mathbf{n}_{2}}M_{\mathbf{n}_{1}\mathbf{n}%
_{2}}(\mathbf{R})\frac{1}{\xi _{1}+\xi _{2}}M_{\mathbf{n}_{1}\mathbf{n}_{2}}(%
\mathbf{R}^{\prime })-\int \frac{d\mathbf{p}}{(2\pi )^{3}}\int \frac{d%
\mathbf{q}}{(2\pi )^{3}}\frac{p_{z}^{2}}{2(p^{2}/2m-\mu )}\exp (i\mathbf{q}(%
\mathbf{R}-\mathbf{R}^{\prime }))\right] \Delta _{0}(\mathbf{R}^{\prime }) 
\notag \\
&\equiv &\int d\mathbf{R}^{\prime }\left[ K_{1}(\mathbf{R},\mathbf{R}%
^{\prime })+K_{2}(\mathbf{R},\mathbf{R}^{\prime })+K_{3}(\mathbf{R},\mathbf{R%
}^{\prime })\right] \Delta _{0}(\mathbf{R}^{\prime }),  \label{gapeqsplit}
\end{eqnarray}%
because the sum $\xi _{1}+\xi _{2}$ does not depend on $n_{1}-n_{2}$ and,
therefore, with the help of formulae (\ref{sumcompltz}) and (\ref%
{sumcompltro}), the summation over $n$ can easily be performed in kernels $%
K_{2}$ and $K_{3}$. On the other hand, the kernel $K_{1}$ is determined
entirely by the states near the chemical potential $\mu $.

The calculation of the sums in the kernel $K_{3}(\mathbf{R},\mathbf{R}%
^{\prime })$ gives

\begin{equation*}
\sum_{\mathbf{n}_{1},\mathbf{n}_{2}}M_{\mathbf{n}_{1}\mathbf{n}_{2}}(\mathbf{%
R})\frac{1}{\xi _{1}+\xi _{2}}M_{\mathbf{n}_{1}\mathbf{n}_{2}}(\mathbf{R}%
^{\prime })=\sum_{\mathbf{N}}\sum\limits_{\mathbf{n}_{1}-\mathbf{n}_{2}}M_{%
\mathbf{Nn}}(\mathbf{R})\frac{1}{2\xi (\mathbf{N})}M_{\mathbf{Nn}}(\mathbf{R}%
^{\prime })
\end{equation*}

\begin{eqnarray*}
&=&\delta (\mathbf{R}-\mathbf{R}^{\prime })\sum_{N_{i}\geq
R_{i}^{2}/2l_{0i}^{2}}\frac{1}{2\xi (\mathbf{N})}\frac{(n\omega _{z})^{2}}{%
\pi ^{3}}\sqrt{\frac{l_{zN}^{2}-z^{2}}{(l_{\rho N}^{2}-x^{2})(l_{\rho
N}^{2}-y^{2})}}=\delta (\mathbf{R}-\mathbf{R}^{\prime })\sum_{N_{i}\geq 0}%
\frac{1}{2\xi (\mathbf{N})}\frac{(n\omega _{z})^{2}}{\pi ^{3}}\frac{l_{zN}}{%
l_{\rho N}l_{\rho N}} \\
&=&\delta (\mathbf{R}-\mathbf{R}^{\prime })\int_{0}\prod_{i}\frac{dp_{i}}{%
\pi }\frac{p_{z}^{2}}{2(p^{2}/2m-\mu (\mathbf{R}))},
\end{eqnarray*}%
where $\mu (\mathbf{R})=\mu -\sum_{i}m\omega _{i}^{2}R_{i}^{2}/2$, and we
have replaced the discrete sums over $N_{i}$ with the integrals over
contineous variables $p_{i}=\sqrt{2m\omega _{i}N_{i}}$. As a result, the
kernel $K_{3}(\mathbf{R},\mathbf{R}^{\prime })$ is

\begin{equation}
K_{3}(\mathbf{R},\mathbf{R}^{\prime })\approx \int \frac{d\mathbf{p}}{(2\pi
)^{3}}\left[ \frac{p_{z}^{2}}{2(p^{2}/2m-\mu (\mathbf{R}))}-\frac{p_{z}^{2}}{%
2(p^{2}/2m-\mu )}\right] \delta (\mathbf{R}-\mathbf{R}^{\prime }).
\label{K3}
\end{equation}%
Following Eq.(\ref{gammaeq}), we see that the kernel $K_{3}$ results in the
replacement $\mu \rightarrow \mu (\mathbf{R})$ in the scattering amplitude $%
\widetilde{\gamma }_{1}(\mu )$ on the left-hand-side of the gap equation (%
\ref{gapeqsplit}).

We are interested in the critical value $\lambda _{c}$ of the aspect ratio,
below which the BCS pairing does not take place for a given strength of the
dipole interaction. Therefore, this value corresponds to vanishing critical
temperature. We therefore calculate the kernels $K_{1}$ and $K_{2}$ in the
limit $T\ll \omega _{i}$. The summation over $\xi (\mathbf{N})=(\xi
_{1})+\xi _{1})/2$ in the kernel $K_{2}$ is then within the limits $-\mu
\leq \xi (\mathbf{N})\leq -\omega _{z}/2$, where $-\omega _{z}/2$ is the
upper limit due to the fact that the function $M_{n_{1}n_{2}}^{(z)}$ is
nonzero only when $\left\vert n_{1z}-n_{2z}\right\vert \geq 1$. These sums
can again be replaced by integrals with the following result

\begin{equation}
K_{2}(\mathbf{R},\mathbf{R}^{\prime })\approx \frac{1}{3}p_{F}^{2}(\mathbf{R}%
)\nu _{F}(\mathbf{R})\left\{ \ln \frac{2\mu (\mathbf{R})}{\omega _{z}}-\frac{%
2}{3}(4-2\ln 2)\right\} \delta (\mathbf{R}-\mathbf{R}^{\prime }),  \label{K2}
\end{equation}
where $p_{F}(\mathbf{R})=\sqrt{2m\mu (\mathbf{R})}$ is the local Fermi
momentum in the Thomas-Fermi approximation.

In order to calculate the kernel $K_{1}$ in the limt $T\ll \omega _{i}$ we
note, that nonzero contributions to the sums over $\mathbf{n}_{1}$ and $%
\mathbf{n}_{2}$ originate only from the region $\omega _{z}/2\leq \left\vert
\xi (\mathbf{N})\right\vert \leq \varepsilon (\mathbf{n})/2$, where $%
n_{i}=\left\vert n_{1i}-n_{2i}\right\vert $ and $\varepsilon (\mathbf{n}%
)=\hbar \sum_{i}\omega _{i}n_{1}$. As a result, the kernel $K_{1}(\mathbf{R},%
\mathbf{R}^{\prime })$ can be written as

\begin{eqnarray*}
K_{1}(\mathbf{R},\mathbf{R}^{\prime }) &=&-\sum_{n_{z}>0;n_{x},n_{y}\geq
0}\delta _{n_{x}}\delta _{n_{y}}\sum_{\omega _{z}/2\leq \xi (\mathbf{N})\leq
\varepsilon (\mathbf{n})/2}\frac{1}{\xi (\mathbf{N})}M_{\mathbf{Nn}}(\mathbf{%
R})M_{\mathbf{Nn}}(\mathbf{R}^{\prime }) \\
&=&-\sum_{n_{z}>0;n_{x},n_{y}\geq 0}\delta _{n_{x}}\delta
_{n_{y}}\int_{\omega _{z}/2}^{\varepsilon (\mathbf{n})/2}\frac{d\xi }{\xi }%
\sum_{\mathbf{N}}\delta (\xi -\xi (\mathbf{N}))M_{\mathbf{Nn}}(\mathbf{R})M_{%
\mathbf{Nn}}(\mathbf{R}^{\prime }).
\end{eqnarray*}

For a generic cylindrical trap, $f(\xi ,\mathbf{n})=\sum_{\mathbf{N}}\delta
(\xi -\xi (\mathbf{N}))M_{\mathbf{Nn}}(\mathbf{R})M_{\mathbf{Nn}}(\mathbf{R}%
^{\prime })$ is a smooth function of $\zeta $ (this is equivalent to
neglecting the so-called shell effect). At the same time, only small $%
\mathbf{n}$ with $\varepsilon (\mathbf{n})\ll \mu $ contributes to the
kernel $K_{1}$. We therefore can replace $f(\xi ,\mathbf{n})$ with $f(0,%
\mathbf{n})$. After replacing the sums over $N_{i}$, $i=x,y,z$, with the
integrals over $\alpha =\omega _{\rho }N_{x}/\mu $, $\beta =\omega _{\rho
}N_{y}/\mu $, and $\zeta =\omega _{z}N_{z}/\mu $, we obtain

\begin{eqnarray}
K_{1}(\mathbf{R},\mathbf{R}^{\prime }) &\approx
&-\sum_{n_{z}>0;n_{x},n_{y}\geq 0}\delta _{n_{x}}\delta _{n_{y}}\ln \frac{%
\varepsilon (\mathbf{n})}{\omega _{z}}\sum_{\mathbf{N}}\delta (\xi (\mathbf{N%
}))M_{\mathbf{Nn}}(\mathbf{R})M_{\mathbf{Nn}}(\mathbf{R}^{\prime })  \notag
\\
&\approx &-\sum_{n_{z}>0;n_{x},n_{y}\geq 0}\delta _{n_{x}}\delta _{n_{y}}%
\frac{\mu ^{2}}{\omega _{\rho }^{2}\omega _{z}}\ln \frac{\varepsilon (%
\mathbf{n})}{\omega _{z}}\int_{0}^{1}d\alpha d\beta d\zeta \delta (1-\alpha
-\beta -\zeta )M_{\mathbf{Nn}}(\mathbf{R})M_{\mathbf{Nn}}(\mathbf{R}^{\prime
}),  \label{K1}
\end{eqnarray}%
where $N_{x}=\alpha \mu /\omega _{\rho }$, $N_{y}=\beta \mu /\omega _{\rho }$%
, and $N_{z}=\zeta \mu /\omega _{z}$.

We write the order parameter in the form $\Delta _{0}(\mathbf{R})=\Delta
_{0}(zR_{TF}^{(z)},xR_{TF}^{(\rho )},yR_{TF}^{(\rho )})=\Delta _{0}(\mathbf{r%
})$, where $R_{TF}^{(i)}=p_{F}/m\omega _{i}$ is the Thomas-Fermi radius of
the gas cloud in the $i$-direction and $\left\vert x\right\vert ,\left\vert
y\right\vert ,\left\vert z\right\vert \leq 1$ are dimensionless variables.
After combinning together Eqs.(\ref{gapeqsplit}),(\ref{K1}),(\ref{K2}), and (%
\ref{K3}), the gap equation in limit $T\ll \omega _{i}$ reads

\begin{equation}
\frac{3}{\Gamma }(1-r^{2})\Delta _{0}(\mathbf{r})=(1-r^{2})^{3/2}\left[ \ln 
\frac{2\mu (\mathbf{r})}{\omega _{z}}-\frac{2}{3}(4-\ln 4)\right] \Delta
_{0}(\mathbf{r})-\frac{3\pi ^{2}}{2}\int_{\mathbf{r}}^{\prime }K(\mathbf{r},%
\mathbf{r}^{\prime })\Delta _{0}(\mathbf{r}^{\prime }),  \label{gapeqfinal}
\end{equation}%
where $\Gamma =\left\vert \gamma _{1}\right\vert \nu _{F}$, $\mu (\mathbf{r}%
)=\mu -V_{\mathrm{trap}}(\mathbf{r})$, and

\begin{eqnarray}
&&K(\mathbf{r},\mathbf{r}^{\prime })=\sum_{n_{z}>0;n_{x},n_{y}\geq 0}\delta
_{n_{x}}\delta _{n_{y}}\ln [n_{z}+\frac{\omega _{\rho }}{\omega _{z}}%
(n_{x}+n_{y})]  \notag \\
&&\times \int_{M_{z}}^{1}d\zeta \int_{M_{x}}^{1}d\alpha
\int_{M_{y}}^{1}d\beta \ \delta (1-\zeta -\alpha -\beta )  \notag \\
&&\times \frac{4}{\pi ^{2}}\frac{\sqrt{(\zeta -z^{2})(\zeta -z^{\prime 2})}}{%
\zeta }\mathrm{U}_{n_{z}-1}(\frac{z}{\sqrt{\zeta }})\mathrm{U}_{n_{z}-1}(%
\frac{z^{\prime }}{\sqrt{\zeta }})  \notag \\
&&\times \frac{4}{\pi ^{2}}\frac{1}{\sqrt{(\alpha -x^{2})(\alpha -x^{\prime
2})}}\mathrm{T}_{n_{x}}(\frac{x}{\sqrt{\alpha }})\mathrm{T}_{n_{x}}(\frac{%
x^{\prime }}{\sqrt{\alpha }})  \notag \\
&&\times \frac{4}{\pi ^{2}}\frac{1}{\sqrt{(\beta -y^{2})(\beta -y^{\prime 2})%
}}\mathrm{T}_{n_{y}}(\frac{y}{\sqrt{\beta }})\mathrm{T}_{n_{y}}(\frac{%
y^{\prime }}{\sqrt{\beta }}),  \label{K}
\end{eqnarray}%
with $M_{s}=\max (s^{2},s^{\prime 2})$ for $s=x,y,z$.

Before solving the above equation numerically, let us analyse the behaviour
of its solutions near the edge of the gas sample, $r\rightarrow 1$. In this
region $\max (x^{2},x^{\prime 2})=x^{2}$, $\max (y^{2},y^{\prime 2})=y^{2}$, 
$\max (z^{2},z^{\prime 2})=z^{2}$, and it is convenient to introduce the new
variables $a=\alpha -x^{2}$, $b=\beta -y^{2}$,\thinspace\ and $c=\zeta
-z^{2} $. The delta function in Eq. (\ref{K}) then reads $\delta
(1-r^{2}-a-b-c)$ and, therefore, only small $a,b,c\leq 1-r^{2}\rightarrow 0$
contribute to the integral. This allows as to write the integral in the last
term in Eq. (\ref{gapeqfinal}) in the form

\begin{eqnarray}
\int d\mathbf{r}^{\prime }K(\mathbf{r},\mathbf{r}^{\prime })\Delta _{0}(%
\mathbf{r}^{\prime }) &\approx &\left( \frac{4}{\pi ^{2}}\right) ^{3}\frac{2%
}{3\pi ^{2}}\sum_{n_{z}>0;n_{x},n_{y}\geq 0}\delta _{n_{x}}\delta
_{n_{y}}\ln [n_{z}+\frac{\omega _{\rho }}{\omega _{z}}(n_{x}+n_{y})]%
\int_{0}^{1-r^{2}}dadbdc\sqrt{\frac{c}{ab}}\delta (1-r^{2}-a-b-c)  \notag \\
&&\ast \int d\mathbf{r}^{\prime }\sqrt{\frac{1-z^{\prime 2}}{(1-x^{\prime
2})(1-y^{\prime 2})}}\mathrm{U}_{n_{z}-1}(1)\mathrm{U}_{n_{z}-1}(z^{\prime })%
\mathrm{T}_{n_{x}}(1)\mathrm{T}_{n_{x}}(x^{\prime })\mathrm{T}_{n_{y}}(1)%
\mathrm{T}_{n_{y}}(y^{\prime })\Delta _{0}(xx^{\prime },yy^{\prime
},zz^{\prime })  \notag \\
&\approx &(1-r^{2})^{3/2}\frac{512}{3\pi ^{5}}\sum_{n_{z}>0;n_{x},n_{y}\geq
0}\delta _{n_{x}}\delta _{n_{y}}\ln [n_{z}+\frac{\omega _{\rho }}{\omega _{z}%
}(n_{x}+n_{y})]\mathrm{U}_{n_{z}-1}(1)\mathrm{T}_{n_{x}}(1)\mathrm{T}%
_{n_{y}}(1)  \notag \\
&&\ast \int d\mathbf{r}^{\prime }\sqrt{\frac{1-z^{\prime 2}}{(1-x^{\prime
2})(1-y^{\prime 2})}}\mathrm{U}_{n_{z}-1}(z^{\prime })\mathrm{T}%
_{n_{x}}(x^{\prime })\mathrm{T}_{n_{y}}(y^{\prime })\Delta _{0}(xx^{\prime
},yy^{\prime },zz^{\prime }),  \label{asympt}
\end{eqnarray}%
and simple analysis of Eq. (\ref{gapeqfinal}) gives the asymptotics $\Delta
_{0}(\mathbf{r})\sim (1-r^{2})^{1/2}$ for $r\rightarrow 1$. Eq. (\ref%
{gapeqfinal}) in the limit $r\rightarrow 0$ and Eq. (\ref{asympt}) can be
thus used as a consistency test for numerical solutions of Eq. (\ref%
{gapeqfinal}).

Our numerical approach to Eq.(\ref{gapeqfinal}) is based on the observation
that it can be considered as the equation on extrema of the quadratic form 
\begin{equation*}
F[\Delta _{0}]=\frac{1}{2}\int_{\mathbf{r},\mathbf{r}^{\prime }}\Delta _{0}(%
\mathbf{r})[L(\mathbf{r})\delta (\mathbf{r}-\mathbf{r}^{\prime })-K(\mathbf{r%
},\mathbf{r}^{\prime })]\Delta _{0}(\mathbf{r}^{\prime })
\end{equation*}%
with $L(\mathbf{r})=3(1-r^{2})/\Gamma -(1-r^{2})^{3/2}\{\ln ([2\mu (\mathbf{r%
})/\omega _{z}]-2(4-\ln 4)/3\}$. Taking into account the asymptotics of the
solutions of Eq.(\ref{gapeqfinal}), we will use the ansatz 
\begin{equation}
\Delta _{0}(\mathbf{r})=(1-r^{2})^{1/2}\sum_{m_{z},m_{\rho }\geq
0}^{M_{z},M_{\rho }}c_{m_{z}m_{\rho }}\mathrm{U}_{m_{z}}(z^{2})\mathrm{T}%
_{m_{\rho }}(x^{2}+y^{2})  \label{ansatz}
\end{equation}%
and write the functional $F[\Delta _{0}]$ as a quadratic form of unknown
coefficients $c_{m_{z}m_{\rho }}$ 
\begin{equation*}
F[c]=\sum_{n_{z},n_{\rho },m_{z},m_{\rho }\geq 0}M_{m_{z},m_{\rho
},n_{z},n_{\rho }}c_{m_{z}m_{\rho }}c_{n_{z}n_{\rho }}
\end{equation*}%
with

\begin{eqnarray}
M_{m_{z},m_{\rho },n_{z},n_{\rho }} &=&\frac{1}{2}\int d\mathbf{r}\left\{ 
\frac{3(1-r^{2})}{\Gamma }-(1-r^{2})^{3/2}\left[ \ln \frac{2\mu }{\omega _{z}%
}(1-r^{2})-\frac{2}{3}(4-2\ln 2)\right] \right\}  \notag \\
&&U_{m_{z}}(z^{2})T_{m_{\rho }}(x^{2}+y^{2})U_{n_{z}}(z^{2})T_{n_{\rho
}}(x^{2}+y^{2})  \notag \\
&&-\frac{3\pi ^{2}}{2}\sum_{n_{z}>0;n_{x},n_{y}\geq 0}\delta _{n_{x}}\delta
_{n_{y}}\ln [n_{z}+\frac{\omega _{\rho }}{\omega _{z}}(n_{x}+n_{y})]  \notag
\\
&&\int_{0}^{1}d\alpha d\beta d\zeta \ \delta (1-\alpha -\beta -\zeta )\zeta
V_{\mathbf{n}}^{m_{z},m_{\rho }}(\alpha ,\beta ,\zeta )V_{\mathbf{n}%
}^{n_{z},n_{\rho }}(\alpha ,\beta ,\zeta ),  \label{M2}
\end{eqnarray}%
where the functions $V$ are defined as: 
\begin{equation}
V_{\mathbf{n}}^{n_{z},n_{\rho }}(\alpha ,\beta ,\zeta )=\int_{-1}^{1}\frac{dx%
}{\pi }\frac{\mathrm{T}_{n_{x}}(x)}{\sqrt{1-x^{2}}}\int_{-1}^{1}\frac{dy}{%
\pi }\frac{\mathrm{T}_{n_{y}}(y)}{\sqrt{1-y^{2}}}\int_{-1}^{1}\frac{dz}{\pi }%
\sqrt{1-z^{2}}\mathrm{U}_{n_{z}-1}(z)U_{n_{z}}(\zeta z^{2})T_{n_{\rho
}}(\alpha x^{2}+\beta y^{2})  \label{MV}
\end{equation}

An extremum of this form at $T_{c}$ corresponds to the eigenvector of the
matrix $M_{m_{z},m_{\rho },n_{z},n_{\rho }}$ with a zero eigenvalue.

The condition that the matrix $M$ has a zero eigenvalue imposes a constraint
on the interaction parameter $\Gamma $ and the trap frequencies $\omega _{i}$
and, therefore, allows to determine the critical aspect ratio. Indeed, the
parameter $\Gamma $ can be written as $\Gamma =36n(0)d^{2}/\pi \mu $, where $%
n(0)=(2m\mu )^{3/2}/6\pi ^{2}\hbar ^{3}$ is the gas density in the center of
the trap, and, hence, for a given dipole moment $d$, $\Gamma $ depends only
on the chemical potential $\mu $. On the other hand, the chemical potential $%
\mu $ and the total number of particles in the trap $N$ are also related, $%
3N=\mu ^{3}/\omega _{z}\omega _{\rho }^{2}$. As a result, for a given $%
\Gamma $ and $N$, the product of the trap frequencies $\omega _{z}\omega
_{\rho }^{2}$ is completely determined, and the only free parameter is the
trap aspect ratio $\lambda =(\omega _{z}/\omega _{\rho })^{1/2}$. We may
thus determine its critical value $\lambda _{c}$ from the condition that the
lowest eigenvalue is zero.

The calculation of the matrix elements $M_{m_{z},m_{\rho },n_{z},n_{\rho }}$
is naturally divided into two parts (see Eq. (\ref{gapeqfinal})). The
contribution with the local kernel $L(\mathbf{r})$ is a three-dimensional
integral that can easily be computed using, for instance, the Monte Carlo
integration routine, such as the VEGAS algorithm from the GSL library ~\cite%
{GSL}. The part with the non-local kernel $K(\mathbf{r},\mathbf{r}^{\prime
}) $ is a triple sum, which elements are eight-dimensional integrals. The
straightforward approach based on the same numerical method fails in this
case because it is too time consuming. To overcome this problem, we
calculate the functions $V_{\mathbf{n}}^{n_{z},n_{\rho }}(\alpha ,\beta
,\zeta )$ in the following way. We integrate numerically over $\mathbf{r}%
^{\prime }$ for fixed $\alpha $, $\beta $ and $n$'s (the value of $\zeta $
is then fixed by the $\delta $-function) and use these data for a
two-dimensional spline interpolation of the integrand for the last
integrations over $\alpha $ and $\beta $. In this way, the time required to
compute the matrix elements $M_{m_{z},m_{\rho },n_{z},n_{\rho }}$ reduces to
about seventy two hours on a standard workstation. This procedure gives
convergent results with respect to the highest powers of the polynomials $%
M_{\zeta }$ and $M_{\rho }$ in our ansatz, Eq. (\ref{ansatz}). We also
checked that our solutions do not depend on the number of shots in the Monte
Carlo algorithm and on the number of points chosen for interpolation. They
are also a proper asymptotic behavior for $r\rightarrow 0$.

The results of our calculations are presented in 2 figures. Fig.2 shows the
desired relation between the interation strength $\Gamma $ and the inverse
critical aspect ratio $\lambda _{c}^{-1}$. The three curves correspond to
three different numbers of particles. As it could be expected, for larger
number of particles, the critical aspect ratio $\lambda _{c}$ is smaller
because the interaction is stronger. For a pancake trap, $\lambda ^{-1}<1$,
the interaction is predominantly repulsive, and higher values of $\Gamma $
for a fixed $\lambda $ are required to achieve the BCS transition. On the
other hand, for a cigar trap, $\lambda ^{-1}>1\,$, the interaction is
predominantly attractive and the BCS transition takes place at smaller
values of $\Gamma $. The existence of the critical interaction strength (for
a given value of $\lambda $) is a result of the discreteness of the spectrum
in the trap and of the specific structure of the order parameter ($\sim
p_{z} $). The latter allows pairing only between particles in the states,
where quantum numbers $n_{z}$ differ by an odd number (intershell pairing).
Therefore, the pairing correlations have to be strong enough to overcome the
corresponding energy difference.

Fig. 3 shows the order parameter $\Delta _{0}(\mathbf{r})$ for the cigar
trap with $\lambda ^{-1}=2.2$. The order parameter is a non-monotonic
function of the distance from the trap center, in contrast to the case of
the BCS order parameter in a two component Fermi gas with short range
interactions ~\cite{monotonicDelta}. This effect persists, although being
less pronounced, for the case of a pancake geometry. In the axial direction,
the order parameter $\Delta (z,\rho =0)$ develops a minimum at $\rho <1$,
whereas in the radial direction $\Delta (z=0,\rho )$ becomes negative in the
outer part of the cloud. This completely new behavior, originating from the
anisotropy of the interpaticle interaction, can have profound consequences
for the form and spectrum of the elementary excitations.

\section{Conclusions}

We have presented a detailed theory and analyzed the influence of a trapping
potential on the BCS pairing in a dipolar single-component Fermi gas. We
have determined the phase diagram for trapped dipolar Fermi gases in the $%
\Gamma -\lambda ^{-1}$\ plane, where $\Gamma $\ measures the dipole strength
and $\lambda $\ is the trap aspect ratio. The BCS transition at finite
temperature $T$\ is possible iff $\Gamma >\Gamma _{c}(\lambda )$. We have
calculated the critical line $\Gamma _{c}(\lambda )$, and the order
parameter at criticality.

\section{Acknowledgements}

We acknowledge support from the DFG, the RTN Cold Quantum Gases, ESF PESC
BEC2000+, the Russian Foundation for Basic Research, QUDEDIS, INTAS, and
from the Alexander von Humboldt Foundation.

\begin{figure}[tbp]
\epsfxsize 7.0cm \centerline{
\epsfbox{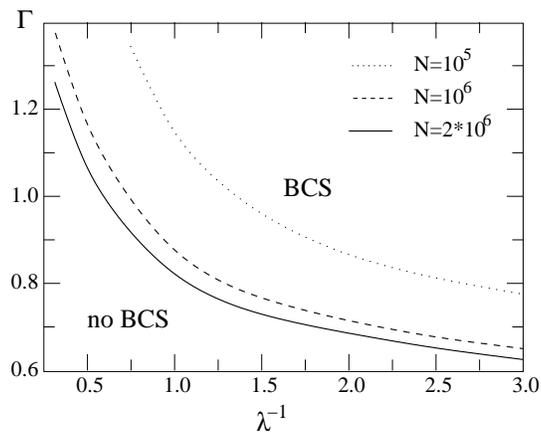}}
\caption{The critical lines $\Gamma _{c}$ versus the inverse aspect ratio $%
\protect\lambda^{-1}$ for different numbers of particles. The BCS pairing
takes place above the depicted curves.}
\label{fig:1}
\end{figure}

\begin{figure}[tbp]
\epsfxsize 6.5cm \centerline{
\epsfbox{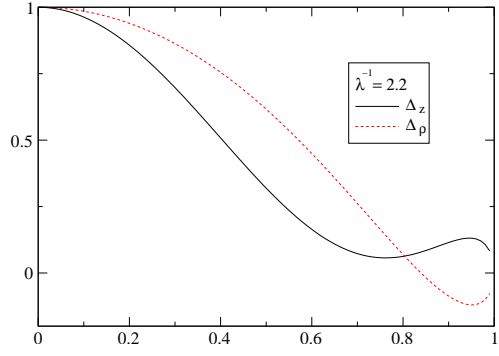}}
\caption{The order parameter for the aspect ratio $\protect\lambda =0.45$
(cigarshape trap). The solid line shows $\Delta _{0}(z,\protect\rho =0)$,
and the dashed line corresponds to $\Delta _{0}({z=0,\protect\rho })$.}
\label{fig:4}
\end{figure}

\section{Appendix}

Here we present the derivation of the WKB-expressions for the functions $%
M_{n_{1}n_{2}}^{(\rho )}(x)$ and $M_{n_{1}n_{2}}^{(z)}(z)$, Eqs. (\ref%
{WKBMro}) and (\ref{WKBMz}). Our starting point is the WKB-eigenfunctions $%
\varphi _{n}(x)$ of a one-dimensional harmonic oscillator with the
Hamiltonian 
\begin{equation*}
H=-\frac{\hbar ^{2}}{2m}\frac{d^{2}}{dx^{2}}+\frac{m\omega ^{2}x^{2}}{2},
\end{equation*}%
with the eigenvalues $E_{n}=\hbar \omega (n+1/2)$,%
\begin{equation}
\varphi _{n}(x)=\sqrt{\frac{2m\omega }{\pi p_{n}(x)}}\cos [\Phi _{n}(x)-\pi
/4],  \label{WKB}
\end{equation}%
where

\begin{equation*}
p_{n}(x)=\sqrt{2m(E_{n}-m\omega ^{2}x^{2}/2)}
\end{equation*}%
and

\begin{equation*}
\Phi _{n}(x)=\frac{1}{\hbar }\int_{-\sqrt{2E_{n}/m\omega }%
}^{x}p_{n}(x^{\prime })dx^{\prime }=\frac{\pi }{2}n+\frac{1}{\hbar }%
\int_{0}^{x}p_{n}(x^{\prime })dx^{\prime }.
\end{equation*}

As it was already mentioned, the functions $M_{n_{1}n_{2}}^{(\rho )}(x)$ and 
$M_{n_{1}n_{2}}^{(z)}(z)$ with $n_{1}$ and $n_{2}$ close to each other (and
both are much larger than unity) are the most important. We therefore
introduce $N=(n_{1}+n_{2})/2$ and $n=n_{1}-n_{2}$, where $n\ll N$, and write
the expression for $M_{n_{1}n_{2}}^{(\rho )}(x)$ as (see Eq. (\ref{MzMro}))

\begin{eqnarray*}
M_{n_{1}n_{2}}^{(\rho )}(x) &\equiv &M_{Nn}^{(\rho )}(x)=\varphi
_{N+n/2}(x)\varphi _{N-n/2}(x) \\
&\approx &\frac{2m\omega }{\pi p_{N}(x)}\frac{1}{2}\left\{ \cos [\Phi
_{N+n/2}(x)-\Phi _{N-n/2}(x)]+\cos [\Phi _{N+n/2}(x)+\Phi _{N-n/2}(x)-\pi
/2]\right\} \\
&\approx &\frac{m\omega }{\pi p_{N}(x)}\left\{ \cos [\Phi _{N+n/2}(x)-\Phi
_{N-n/2}(x)]\right\} ,
\end{eqnarray*}%
where we neglect the rapidly oscillating contribution $\cos [\Phi
_{N+n/2}(x)+\Phi _{N-n/2}(x)-\pi ]$. In the case $n\ll N$, the phase
difference $\Phi _{N+n/2}(x)-\Phi _{N-n/2}(x)$ can be simplified as

\begin{eqnarray}
\Phi _{N+n/2}(x)-\Phi _{N-n/2}(x) &=&\frac{\pi }{2}n+\frac{1}{\hbar }%
\int_{0}^{x}[p_{N+n/2}(x^{\prime })-p_{N-n/2}(x^{\prime })]dx^{\prime } 
\notag \\
&\approx &\frac{\pi }{2}n+n\omega \int_{0}^{x}\frac{\partial p_{N}(x^{\prime
})}{\partial E_{N}}dx^{\prime }  \notag \\
&=&n\left[ \pi -\arccos \left( x\sqrt{2Nm\omega /\hbar }\right) \right] .
\label{phasedifference}
\end{eqnarray}%
As a result, we obtain the expression

\begin{eqnarray*}
M_{n_{1}n_{2}}^{(\rho )}(x) &\approx &(-1)^{n}\frac{m\omega }{\pi p_{N}(x)}%
\cos \left[ n\arccos \left( x\sqrt{2Nm\omega /\hbar }\right) \right] \\
&=&(-1)^{n}\mathrm{T}_{n}\left( x\sqrt{2Nm\omega /\hbar }\right) ,
\end{eqnarray*}%
which corresponds to Eq. (\ref{WKBMro}) in the dimensional units.

The expression (\ref{MzMro}) for the function $M_{n_{1}n_{2}}^{(z)}(z)$
involves the first derivative of the wave function. According to the general
rules of the WKB approximation (see, e.g., \cite{LandauQM}), this derivative
should act only on the rapidly varying trigonometric factor in Eq. (\ref{WKB}%
). Therefore, $M_{n_{1}n_{2}}^{(z)}(z)$ can be written in the form

\begin{eqnarray*}
M_{n_{1}n_{2}}^{(z)}(z) &\equiv &M_{Nn}^{(z)}(z)=\frac{1}{2}\left[ \varphi
_{N+n/2}(z)\partial _{z}\varphi _{N-n/2}(z)-\varphi _{N-n/2}(z)\partial
_{z}\varphi _{N+n/2}(z)\right] \\
&\approx &-\frac{m\omega }{\pi \hbar }\left\{ \sin [\Phi _{N+n/2}(z)-\Phi
_{N-n/2}(z)]\right\} ,
\end{eqnarray*}%
where again we neglect the rapidly oscillating contribution. The application
of Eq. (\ref{phasedifference}) gives

\begin{equation*}
M_{n_{1}n_{2}}^{(z)}(z)\approx (-1)^{n}\frac{m\omega }{\pi \hbar }\sin \left[
n\arccos \left( x\sqrt{2Nm\omega /\hbar }\right) \right] ,
\end{equation*}%
and, assuming $n>0$, we obtain

\begin{equation*}
M_{n_{1}n_{2}}^{(z)}(z)\approx (-1)^{n}\frac{m\omega }{\pi \hbar }\sqrt{%
1-\left( x\sqrt{2Nm\omega /\hbar }\right) ^{2}}\mathrm{U}_{n-1}\left( x\sqrt{%
2Nm\omega /\hbar }\right) .
\end{equation*}%
Note also, that one obtains the same result if, instead of differentiating
the WKB wave function, Eq. (\ref{WKB}), one uses the well-known relations
between the wave functions of a harmonic oscillator and its derivatives.



\end{document}